\documentclass[12pt]{article}
\usepackage{a4wide}
\usepackage{amssymb}
\usepackage{graphicx}
\begin{document}
{\renewcommand{\thefootnote}{\fnsymbol{footnote}}
\hfill CGPG--01/5--1 \\
\medskip
\hfill gr--qc/0105067\\
\medskip
\begin{center}
{\LARGE  The Inverse Scale Factor\\[0.5em] in Isotropic Quantum Geometry}\\
\vspace{1.5em}
Martin Bojowald\footnote{e-mail address: {\tt bojowald@gravity.phys.psu.edu}}\\
\vspace{0.5em}
Center for Gravitational Physics and Geometry,\\
The Pennsylvania State
University,\\
104 Davey Lab, University Park, PA 16802, USA
\vspace{1.5em}
\end{center}
}

\setcounter{footnote}{0}

\newcommand{\case}[2]{{\textstyle \frac{#1}{#2}}}
\newcommand{\lP}{l_{\mathrm P}}
\newcommand{\md}{{\mathrm{d}}}
\newcommand{\tr}{\mathop{\mathrm{tr}}}
\newcommand{\sgn}{\mathop{\mathrm{sgn}}}

\newcommand*{\R}{{\mathbb R}}
\newcommand*{\Z}{{\mathbb Z}}

\begin{abstract}
  The inverse scale factor, which in classical cosmological models
  diverges at the singularity, is quantized in isotropic models of
  loop quantum cosmology by using techniques which have been developed
  in quantum geometry for a quantization of general relativity.  This
  procedure results in a {\em bounded\/} operator which is
  diagonalizable simultaneously with the volume operator and whose
  eigenvalues are determined explicitly. For large scale factors (in
  fact, up to a scale factor slightly above the Planck length) the
  eigenvalues are close to the classical expectation, whereas below
  the Planck length there are large deviations leading to a
  non-diverging behavior of the inverse scale factor even though the
  scale factor has vanishing eigenvalues.  This is a first indication
  that the classical singularity is better behaved in loop quantum
  cosmology.
\end{abstract}

\section{Introduction}

General relativity predicts singularities in many situations of
astrophysical or cosmological interest, which means that there are
limits for the classical theory beyond which it is no longer valid. A
widespread expectation is that a quantization of gravity is inevitable
in order to describe these regimes meaningfully, but up to now there
is no complete, generally accepted quantization of general relativity.
For a long time, mini-superspace models obtained by a symmetry
reduction of the classical theory with a subsequent quantization
\cite{DeWitt,Misner} (henceforth called ``standard quantum cosmology''
in the context of cosmological models) were the only approach to
address those issues; but in view of the fact that the quantization
techniques were those of simple quantum mechanical systems, which
cannot be applied to the full theory, the results are not likely to
hold true in a full quantization. In fact, it has been shown in
quantum geometry \cite{Rov:Loops} that geometry has a discrete
structure leading, e.g., to a discrete volume spectrum
\cite{AreaVol,Vol2}, whereas in standard quantum cosmology the scale
factor, and so the volume, is still continuous with a range from zero
to infinity. For large volume, this is a very good approximation to
the discrete volume spectrum of quantum geometry, but just in the
domain close to the classical singularity there are huge deviations
between the discrete and the continuous spectra.

Therefore, we follow a different approach to quantum cosmology which
has been initiated in
\cite{SymmRed,cosmoI,cosmoII,cosmoIII,cosmoIV,PhD} and which starts by
selecting symmetric (isotropic, and in particular homogeneous, in this
paper) states in the kinematical Hilbert space of quantum
geometry. This means that the symmetry reduction is not purely
classical, but is done after an essential step of the quantization
which already leads to the discrete structure of space. Consequently,
the volume spectrum of cosmological models is discrete and even known
explicitly in the isotropic case \cite{cosmoII}. The simplification of
the spectrum caused by symmetry (interpreted analogously to the
familiar level splitting in the spectroscopy of atoms) is very
fortunate because it facilitates explicit calculations. Reciprocally,
the discreteness of the volume spectrum implies that the models of
loop quantum cosmology embody the distinctive feature of quantum
geometry of having a discrete structure. In fact, the quantization
techniques of loop quantum cosmology are designed to be as close to
those of the full theory as possible, with only slight adaptations to
the symmetry. So those models can be used for crucial tests of methods
developed in the full theory, but they have also been used to derive
new properties, e.g., a discrete time and discrete physical (not just
kinematical) spectra of geometric operators \cite{cosmoIV}.

In this paper we will use techniques which have been developed in
order to quantize the Hamiltonian constraint and matter Hamiltonians
in the full theory \cite{QSDI,QSDV} for a quantization of the inverse
scale factor in isotropic quantum geometry. As result we will derive a
bounded operator despite of the fact that the volume or the scale
factor has vanishing eigenvalues. The underlying ``mechanism'' is the
same as the one which ensures, in the full theory, that matter
Hamiltonians can be quantized to obtain densely defined operators. One
might suspect that this is simply a mathematical trick which serves to
remove singularities but which will spoil the classical limit. We will
show that this is not the case: singularities are removed, but the
classical regime is not affected. In fact, for the inverse scale
factor the classical theory turns out to be an excellent approximation
right up to a scale factor of the order of the Planck length. This is
quite unexpected; a priori, one expects the classical behavior to be
valid only for scale factors very large compared to the Planck length.

The plan of the paper is as follows: First, we will recall the
framework of isotropic loop quantum cosmology and extend the methods
developed in \cite{cosmoII} to gauge non-invariant states in Section
\ref{s:Iso}. This will then be applied for a discussion of the inverse
scale factor which is quantized in Section \ref{s:Inv}. We will
determine all its eigenstates and its complete spectrum and study the
two interesting regimes for very small and large scale factors. In a
last section we will present our conclusions concerning the validity
of quantization techniques and the quantum picture of the classical
singularity.

\section{Isotropic Quantum Geometry}
\label{s:Iso}

The general framework for a symmetry reduction of quantized
diffeomorphism invariant theories has been developed in \cite{SymmRed}
and specialized to homogeneous and isotropic models in \cite{cosmoI}.
Symmetric states are defined at the kinematical level of the quantum
theory, and thus have properties very different from those of states
obtained after quantizing a classically reduced theory. Still, for
explicit expressions of symmetric states and operators we need to know
the symmetry reduction of a theory of connections and triads which we
will sketch first in the case of isotropy.

Isotropic connections are of the form $A_a^i= \phi_I^i \omega_a^I= c
\Lambda_I^i \omega_a^I$, where $\Lambda_I=\Lambda_I^i\tau_i$ is an
internal $SU(2)$-dreibein (which is purely a gauge choice) and
$\omega^I$ are left-invariant one-forms on the ``translational part''
$N$ of the symmetry group $S\cong N\rtimes SO(3)$ acting on the space
manifold $\Sigma$. (Here, $\tau_j= -\frac{i}{2} \sigma_j$ are
generators of $SU(2)$ with $\sigma_j$ the Pauli matrices; $N$ is
isomorphic to $\R^3$ for the spatially flat model or $SU(2)$ for the
spatially positively curved model.) For homogeneous models, the nine
parameters $\phi_I^i$ are arbitrary. A co-triad can be expressed as
$e_a^i=a_I^i \omega_a^I= a \Lambda_I^i \omega_a^I$ with the scale
factor\footnote{In a triad formulation we use a variable $a$ which can
take both signs, even though the two sectors of positive and negative
$a$, respectively, are disconnected in a metric formulation.}
$|a|$. Using left-invariant vector fields $X_I$ fulfilling
$\omega^I(X_J)=\delta^I_J$, momenta canonically conjugate to $A_a^i$
are densitized triads of the form\footnote{Note that, in contrast to
\cite{cosmoI}, we use the physical metric in order to provide the
density weight and not an auxiliary homogeneous metric: $p^I_i:=|\det
a^j_J| a_i^I$, $a^I_i$ being inverse to $a_I^i$. Nevertheless, we need
to fix a reference system already in order to define the action of our
symmetry group, which leads to the factor of $V_0$ in the formulae of
\cite{cosmoI,cosmoII}. However, this factor is an artifact of the
homogeneous models and not of physical significance.  It just tells us
that we cannot define an absolute scale factor in a diffeomorphism
invariant setting, but only a relative one with respect to some given
value. In this paper we will set $V_0=1$.} $E_i^a= p_i^I X_I^a= p
\Lambda_i^I X_I^a$ where $p=\sgn(a) a^2$. Besides gauge freedom, there
are only the two canonically conjugate variables $c$ and $p$ which
embody the gauge invariant information of the connection and
triad. Information about the geometry of space is fully contained in
$p$, which is the square of the radius of a spacelike slice. The
Liouville form
\[
 (\gamma\kappa)^{-1}
 p_i^I \md\phi_I^i= (\gamma\kappa)^{-1} p\md c\Lambda_i^I \Lambda_I^i=
 3(\gamma\kappa)^{-1} p\md c
\]
leads to the symplectic structure
\begin{equation}\label{symp}
 \{c,p\}=\case{1}{3}\gamma\kappa
\end{equation}
($\kappa=8\pi G$ is the gravitational constant and $\gamma$ the
Barbero--Immirzi parameter).  The factor $\frac{1}{3}$ has been
overlooked in \cite{cosmoII}, and so also the derivative operators and
the volume spectrum derived there have to be corrected by appropriate
factors. We will do this in the formulae below.

Isotropic states in the connection representation are defined as
distributional states in the full theory which are supported only on
isotropic connections. Since all the information of an isotropic
connection is contained in one $SU(2)$-element
($\exp(c\Lambda_3^i\tau_i)$, say) the reduced kinematical Hilbert
space can be taken to be the space ${\cal H}_{\rm kin}=L^2(SU(2),{\rm
d}\mu_H)$ of square integrable functions on $SU(2)$ with respect to
Haar measure. However, gauge invariance is not imposed in the obvious
sense by conjugation on this copy of $SU(2)$, but instead in such a
way that there is a larger class of gauge invariant functions (see
\cite{cosmoII} for details).  Besides the usual character functions
\begin{equation}\label{chi}
 \chi_j=\frac{\sin(j+\case{1}{2})c}{\sin\case{1}{2}c}\,,
\end{equation}
where $j$ is a non-negative half-integer, we have gauge-invariant
states given by $\zeta_{-\frac{1}{2}}= (\sqrt{2}
\sin\case{1}{2}c)^{-1}$ and
\begin{equation}\label{zeta}
 \zeta_j=\frac{\cos(j+\case{1}{2})c}{\sin\case{1}{2}c}\,.
\end{equation}
All states $(\chi_j,\zeta_j)$ form an orthonormal basis of the gauge
invariant kinematical Hilbert space. The fact that we need the
functions $\zeta_j$ and also the value $j=-\frac{1}{2}$ can be
seen from a representation of the kinematical Hilbert space as
periodic functions in $c$, where the measure provides a factor
$\sin^2\frac{1}{2}c$. A complete set of those functions is given by
$\sin$ and $\cos$ with the above frequencies. Gauge non-invariant
functions are given by $\Lambda^i_I\chi_j$ and $\Lambda_I^i\zeta_j$
where $\Lambda_I^i$ is the internal dreibein and provides pure gauge
degrees of freedom.

\begin{figure}[ht]
\begin{center}
 \includegraphics[width=12cm,height=8cm,keepaspectratio]{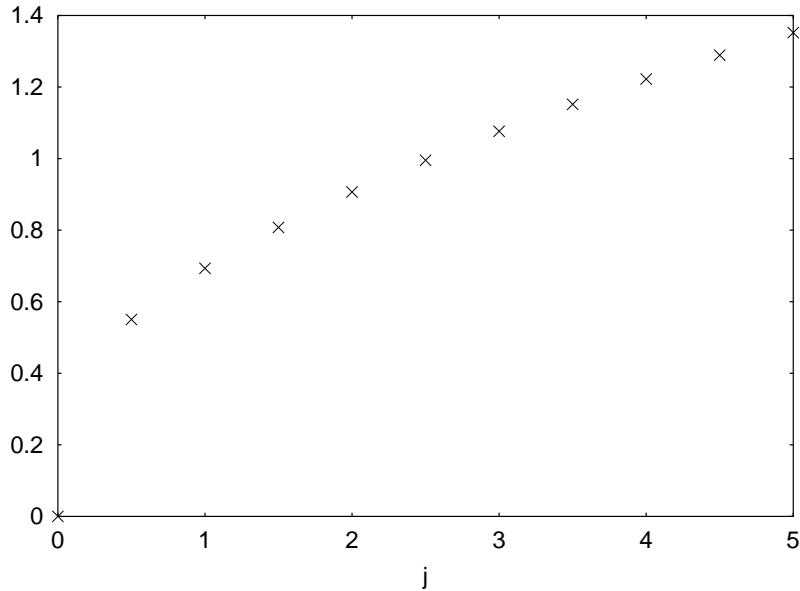}
\end{center}
\caption{Eigenvalues $V_j^{\frac{1}{3}}$
of the scale factor (in units of $\sqrt{\gamma}\lP$).}
\label{scale}
\end{figure}

In \cite{cosmoII} the volume operator $\hat{V}$ has been shown to have
the eigenstates $\chi_j$, $\zeta_j$ with eigenvalues (corrected here
for the missing factor $\frac{1}{3}$ in the symplectic structure and
derivative operators)
\begin{equation}\label{vol}
 V_j=(\gamma\lP^2)^{\frac{3}{2}}\sqrt{\case{1}{27}j(j+\case{1}{2})(j+1)}\,.
\end{equation}
The eigenvalue zero is three-fold degenerate, whereas all other
eigenvalues are positive and twice degenerate. The two-fold degeneracy
arises naturally in a triad formulation because any value of the
volume can be achieved in two different orientations of the
triad. Intuitively, this demonstrates the necessity of the states
$\zeta_j$ besides the characters $\chi_j$.  Taking the cubic root
yields the eigenvalues of the scale factor $|a|$ which are shown in
Fig.\ \ref{scale}

However, in \cite{cosmoII} the action action of $\hat{V}$ has not been
determined on gauge non-invariant states.  An extension to those
states is done by using gauge invariance of the volume, which implies
$[\Lambda^i_I, \hat{V}]= \alpha \Lambda_I^i$ for some
$\alpha\in\R$. We can now use $\sum_i\Lambda_I^i \Lambda_i^I=1$ (no
sum over $I$) in order to obtain
\[
 0=\left[\sum_i\Lambda_I^i \Lambda_i^I, \hat{V}\right]= \sum_i\Lambda_I^i
 [\Lambda_i^I, \hat{V}]+ \sum_i [\Lambda_I^i, \hat{V}] \Lambda_i^I=
 2\alpha
\]
and so $\alpha=0$. Thus, the volume operator commutes with
$\Lambda^i_I$ (acting as multiplication operator in the connection
representation) and we can trivially extend its action to gauge
non-invariant states.

All we need for the following calculations is the action of
$\cos\frac{1}{2}c$ and $\sin\frac{1}{2}c$ appearing in the ``point
holonomy'' \cite{FermionHiggs} $h_I:= \exp(c\Lambda_I^i\tau_i)=
\cos(\frac{1}{2}c)+ 2\sin(\frac{1}{2}c) \Lambda_I^i\tau_i$ which in
quantum geometry serves as the basic multiplication operator. This can
be obtained in the connection representation (\ref{chi}), (\ref{zeta})
by using trigonometric relations leading for $j\geq\frac{1}{2}$ to
\begin{eqnarray}
 \cos(\case{1}{2}c)\:\chi_j = \case{1}{2}(\chi_{j+\frac{1}{2}}+
 \chi_{j-\frac{1}{2}}) 
 \quad &,& \quad \cos(\case{1}{2}c)\:\zeta_j=
 \case{1}{2} (\zeta_{j+\frac{1}{2}}+ \zeta_{j-\frac{1}{2}})\label{cos}\\
 \sin(\case{1}{2}c)\:\chi_j = -\case{1}{2}(\zeta_{j+\frac{1}{2}}-
 \zeta_{j-\frac{1}{2}})
 \quad &,& \quad\sin(\case{1}{2}c)\:\zeta_j=
 \case{1}{2} (\chi_{j+\frac{1}{2}}- \chi_{j-\frac{1}{2}})\label{sin}
\end{eqnarray}
with certain modifications in numerical coefficients for $j=0$ or
$j=-\frac{1}{2}$, which are not important for our purposes.

Because of the exceptional formulae for low $j$ it is more convenient
to use the states
\begin{equation}
 |n\rangle:=\frac{\exp(\case{1}{2}inc)}{\sqrt{2}\sin\case{1}{2}c}
 \quad,\quad n\in\Z
\end{equation}
which are decomposed in the previous states by
\[
 |n\rangle= 2^{-\frac{1}{2}} \left(\zeta_{\frac{1}{2}(|n|-1)}+ i\sgn(n)
 \chi_{\frac{1}{2}(|n|-1)}\right)
\]
for $n\not=0$ and $|0\rangle=\zeta_{-\frac{1}{2}}$. The label $n$ of a
state $|n\rangle$ is proportional to the eigenvalue of the dreibein
operator \cite{cosmoII} (note the factor $\frac{1}{3}$ in order to
correct for the symplectic structure (\ref{symp}))
\[
 \hat{p}= \widehat{\Lambda_3^iE_i^3}= -\case{1}{3}i\gamma\lP^2
 \left(\frac{\md}{\md c}+ \case{1}{2} \cot\case{1}{2}c\right)\,.
\]
On these states the action of $\cos\frac{1}{2}c$ and $\sin\frac{1}{2}c$
is simply
\[
 \cos(\case{1}{2}c)\:|n\rangle= \case{1}{2}(|n+1\rangle+|n-1\rangle)
 \quad,\quad \sin(\case{1}{2}c)\:|n\rangle= -\case{1}{2}i(|n+1\rangle-
 |n-1\rangle)
\]
for all integer $n$.

\section{Quantization of the Inverse Scale Factor}
\label{s:Inv}

In isotropic geometries, the classical singularity is signaled by the
inverse scale factor $|a|^{-1}$ which diverges for $a=0$ and occurs by
some positive power in all diverging curvature
components. Classically, we have $|a|^{-1}=V^{-\frac{1}{3}}$ so that
one might try to quantize it by using the inverse volume
operator. This inverse, however fails to be a densely defined operator
because $\hat{V}$ has the eigenvalue zero (with threefold degeneracy:
$\hat{V}$ annihilates $\zeta_{-\frac{1}{2}}$, $\chi_0$ and $\zeta_0$).
Thus, we have to look for another approach. We will use an expression
which classically reduces to the inverse scale factor but is better
suited for a quantization, namely
\begin{equation}\label{m}
 m_{IJ}:=\frac{q_{IJ}}{\sqrt{\det q}}=
 \frac{a_I^ia_I^i}{|\det (a^i_I)|}= \frac{1}{|a|}\delta_{IJ}
\end{equation}
in terms of the isotropic metric $q_{IJ}=a^2\delta_{IJ}$ or the triad
components $a_I^i$. Since the latter are not fundamental variables,
one needs a prescription to quantize them. Here one can make use of
the classical identity
\begin{equation}\label{trick}
 e_a^i=2(\gamma\kappa)^{-1}\{A_a^i,V\}
\end{equation}
and quantize the co-triad by expressing the connection in terms of a
holonomy, using the volume operator and turning the Poisson bracket
into a commutator. This method has been successfully employed in
\cite{QSDI} in order to quantize the Hamiltonian constraint of the
full theory. In this Section we carry out a similar procedure for a
quantization of the inverse scale factor. The regularization scheme
adapted to isotropic models is reviewed in Section \ref{s:Reg} and
then applied to $m_{IJ}$ in Section \ref{s:Quant} where we derive an
operator $\hat{m}_{IJ}$. Its Spectrum is determined in Section
\ref{s:Spec} followed by a discussion of its main features and
viability (Section \ref{s:Half}).

\subsection{The Regularization Scheme}
\label{s:Reg}

Let us first recall from \cite{cosmoIII} the regularization scheme. As
noted in \cite{QSDI}, it is important to be aware of the density
weight when regularizing expressions in a diffeomorphism invariant
field theory: only scalar quantities, usually space integrals of
weight one densities, can be quantized in a background independent
manner. This is also important in our reduced models. Although we do
not have the freedom to make arbitrary coordinate transformations
since most of them would violate the symmetry conditions, we do have
dilatations with a scale parameter $\epsilon$ which allow us to keep
track of the density weight. These scale transformations also play an
important role in adapting the regularization of \cite{QSDI} in the
full theory to reduced models (see \cite{cosmoIII} for details). We
will proceed along the lines of the following recipe: Starting from a
classical expression in the full theory we first insert homogeneous
fields parameterized by the components $\phi_I^i$, $p_i^I$, \ldots\
and arrive at the reduced expression for a homogeneous model. Now we
start the regularization by performing a scale transformation with
parameter $\epsilon$. Taking care of the density weights, we have to
multiply any one-form component by $\epsilon$, any density-weighted
vector field component by $\epsilon^2$, Poisson brackets by
$\epsilon^{-3}$, etc. Our original expression, a density integrated
over space, then gets multiplied by a factor $\epsilon^3$ which is
absorbed in the rescaled space volume. In a homogeneous model, the
continuum limit needed in the regularization is replaced by a limit
$\epsilon\to 0$ which means, e.g., that connection components
$\phi_I^i$ can be approximated by holonomies $h_I=
\exp(\epsilon\phi_I^i\tau_i)= 1+ \epsilon\phi_I^i\tau_i+
O(\epsilon^2)$ as in the full theory. The quantized expression, on the
other hand, will be independent of $\epsilon$; for a detailed
discussion see \cite{cosmoIII}.

If we are interested in isotropic models, we have to perform another
step because the homogeneous coefficients have to be put in isotropic
form thereby yielding the isotropic classical expressions. In the
quantization, we also start from the quantized homogeneous operator
and insert special holonomies $h_I= \cos(\frac{1}{2}c)+
2\sin(\frac{1}{2}c) \Lambda_I^i\tau_i$ which leads to the operators
$\cos\frac{1}{2}c$ and $\sin\frac{1}{2}c$ whose action we already
know. Derivative operators are treated similarly, but we usually only
need the volume operator which has already been derived
\cite{cosmoII}.

We illustrate a typical calculation by computing the Poisson bracket
\[
 \{\sin\case{1}{2}c,V\}= \{\sin\case{1}{2}c,|p|^{\frac{3}{2}}\}=
 \case{1}{4} \gamma\kappa \cos(\case{1}{2}c) \sqrt{|p|}\sgn(p)
\]
using the symplectic structure (\ref{symp}). The corresponding
commutator of the quantized objects acts on $\chi_j$ as
\begin{eqnarray*}
 [\sin\case{1}{2}c,\hat{V}]\chi_j &=& \case{1}{2} (V_{j+\frac{1}{2}}-V_j)
  \zeta_{j+\frac{1}{2}}+ \case{1}{2} (V_j-V_{j-\frac{1}{2}})
  \zeta_{j-\frac{1}{2}}\\
 &\sim& \case{1}{4\sqrt{3}} (\gamma\lP^2)^{\frac{3}{2}} \sqrt{j}\cdot
 \case{1}{2} (\zeta_{j+\frac{1}{2}}+\zeta_{j-\frac{1}{2}})
\end{eqnarray*}
which we expanded in the last line for large $j$. A quantization of
$\sqrt{|p|}\sgn(p)$ should have the asymptotic behavior $\chi_j\mapsto
i\sqrt{\gamma\lP^2}\sqrt{j/3}\zeta_j$ for large $j$ (because of the
sign it maps $\chi$ to $i\zeta$ and vice versa, and the factor
$\sqrt{j/3}$ follows from the large-$j$ behavior of the volume
spectrum). So we see that for large $j$ (where the ordering is
irrelevant) the commutator is $\frac{1}{4}i \gamma\lP^2
\cos(\frac{1}{2}c) \sqrt{|\hat{p}|}\sgn(\hat{p})$ which demonstrates that
we have the correct expression corresponding to $i\hbar$ times the
classical Poisson bracket. For the correct prefactor $\frac{1}{4}$ it
is important that we used the symplectic structure (\ref{symp}) and
corrected the volume eigenvalues.

\subsection{Quantization}
\label{s:Quant}

In order to quantize the inverse scale factor we use the expression
(\ref{m}). However, it is not a density and thus for a regularization
along the above lines we need to first transform it into an expression
which is a density such that it can be integrated to a scalar. This
can easily be achieved by contracting with two density weighted vector
fields, e.g.\ the electric field $E^a$, leading to the electric part
of the Maxwell Hamiltonian
\[
 H=\int\md^3x\frac{q_{ab}}{\sqrt{\det q}}E^aE^b\,.
\]
In this paper we are only interested in the gravitational part which
will be separated later, but the full expression can be used for
studying, e.g., Maxwell theory coupled to quantum gravity.

First we have to insert homogeneous co-triad components $e_a^i=a_I^i
\omega_a^I$ and also homogeneous electric fields $E^a=E^I X_I^a$ (we
also integrate over space and suppress the resulting factor of the
coordinate volume setting $V_0=1$):
\[
 H=\frac{a_I^ia_J^i}{\sqrt{|\det (p^I_i)|}} E^IE^J\,.
\]
Now we have to express the co-triad components $a_I^i$ by a Poisson
bracket using (\ref{trick}). This expression can be derived by first
computing
\begin{eqnarray*}
 \{\phi_K^k,\epsilon_{MNL}\epsilon^{ijl}p_i^Mp_j^Np_l^L\} &=&
 3\gamma\kappa \epsilon^{ijk}\epsilon_{MNK} p_i^Mp_j^N= 3\gamma\kappa
   \epsilon^{ijk}\epsilon_{ijl} a_K^l \sgn(\det( p_m^M))
 \sqrt{|\det(p_m^M)|}\\
 &=& 6\gamma\kappa
    a_K^k \sgn(\det(p_m^M))\sqrt{|\det(p_m^M)|}
\end{eqnarray*}
where we have used $\delta^L_K= p^L_la^l_K |\det
p_m^M|^{-\frac{1}{2}}$ in the second step. We thus have the reduction
of Thiemann's identity \cite{QSDI} to homogeneous variables:
\begin{eqnarray}\label{tricka}
 a_K^k &=& (\gamma\kappa
 \sgn(\det(p_m^M))\sqrt{|\det(p_m^M)|})^{-1}\{\phi_K^k, \det(p_m^M)\}=
 2(\gamma\kappa)^{-1}\{\phi_K^k, \sqrt{|\det(p_m^M)|}\}\nonumber\\
 &=& 2(\gamma\kappa)^{-1} \{\phi_K^k,V\}\,.
\end{eqnarray}
We insert this expression in $H$ to obtain
\begin{eqnarray*}
 H &=& 4(\gamma\kappa)^{-2}\frac{\{\phi_I^i,V\} \{\phi_J^i,V\}}{V}
  E^IE^J\\
 &=& 16(\gamma\kappa)^{-2} \{\phi_I^i,\sqrt{V}\} \{\phi_J^i,\sqrt{V}\}
 E^IE^J
\end{eqnarray*}
which we now use for the regularization. Note that we were
able to absorb the $V$ in the denominator into the Poisson bracket in
the numerator, as first done in \cite{QSDI}. {\em This is the key point
leading to a bounded operator after quantization}.

We now multiply the components $\phi_I^i$, $E^I$, the volume and the
Poisson brackets by the factors $\epsilon$, $\epsilon^2$, $\epsilon^3$
and $\epsilon^{-3}$, respectively, and obtain the regularized
expression (absorbing $\epsilon^{-3}$ in the space integral)
\begin{eqnarray*}
 H_{\epsilon} &=& 16(\gamma\kappa)^{-2} \{\epsilon
 \phi_I^i,\sqrt{V}\} \{\epsilon\phi_J^i,\sqrt{V}\} \epsilon^2E^I
 \epsilon^2E^J\\
 &=& -32(\gamma\kappa)^{-2} \tr\left(h_I\{h_I^{-1},\sqrt{V}\}
 h_J\{h_J^{-1},\sqrt{V}\}\right) \epsilon^2E^I \epsilon^2E^J+ O(\epsilon^7)\,.
\end{eqnarray*}
Now we can read off the gravitational part and immediately quantize:
The remaining factors of $\epsilon^2$ are needed for a quantization of
the electric field components, whereas the rest yields
\begin{equation}\label{mhom}
 \hat{m}_{IJ}=32 (\gamma\lP^2)^{-2} \tr\left(h_I\left[h_I^{-1},
 \sqrt{\hat{V}}\right] h_J\left[h_J^{-1},\sqrt{\hat{V}}\right]\right)
\end{equation}
which is independent of the regulator $\epsilon$.

So far, we have only used homogeneity; next, we can reduce
(\ref{mhom}) to isotropy: We have to insert the special form of
holonomies and the isotropic volume operator and can then take the
trace over the dreibein components $\Lambda_I$ to arrive at the
isotropic inverse scale factor
\begin{eqnarray}
 \hat{m}_{IJ} &=& 64 (\gamma\lP^2)^{-2} \left(\left( \sqrt{\hat{V}}
 - \cos(\case{1}{2}c)\sqrt{\hat{V}} \cos(\case{1}{2}c) -\sin(\case{1}{2}c)
 \sqrt{\hat{V}} \sin(\case{1}{2}c)\right)^2\right.\nonumber\\
 & & -\delta_{IJ} \left.\left(\sin(\case{1}{2}c) \sqrt{\hat{V}}
   \cos(\case{1}{2}c)-\cos(\case{1}{2}c) \sqrt{\hat{V}}
   \sin(\case{1}{2}c)\right)^2\right)\,.\label{miso}
\end{eqnarray}

This operator has two striking features: First, it provides a
quantization of the inverse scale factor by a {\em bounded operator\/}
despite of the fact that the classical expression diverges for
$a\to0$. As seen in Fig.\ \ref{inv} the upper bound is given by its
eigenvalue in the state with $j=\frac{1}{2}$ and has the value
$\frac{32}{3}\cdot (2-\sqrt{2})(\gamma\lP^2)^{-\frac{1}{2}}$ which
diverges for $\hbar\to0$. Thus it is the finiteness of Planck's
constant that removes the infinity of the classical inverse scale
factor. This is somewhat analogous to the ground state energy of the
hydrogen atom: it is negative and finite in quantum theory, but
diverges for $\hbar\to0$ in correspondence with the fact that the
classical energy is unbounded from below.  The second feature is that
the operator-valued matrix $\hat{m}_{IJ}$ is not diagonal as in the
classical case. As one would expect, the off-diagonal components have
a purely quantum origin and go to zero as $\hbar$ tends to zero. In
order to critically examine the viability of the quantization
procedure we need to discuss the classical limit which we will do
below after deriving the complete spectrum of $\hat{m}_{IJ}$.

\subsection{Spectrum of the Inverse Scale Factor}
\label{s:Spec}

Both squared brackets in the operator (\ref{miso}) act diagonally on
the states $\chi_j$, $\zeta_j$ which can be derived by using the
volume eigenvalues (\ref{vol}) and the operators $\cos\frac{1}{2}c$ and
$\sin\frac{1}{2}c$ in (\ref{cos}), (\ref{sin}). The result is
\begin{eqnarray*}
 \left( \sqrt{\hat{V}} - \cos(\case{1}{2}c)\sqrt{\hat{V}} \cos(\case{1}{2}c)
   -\sin(\case{1}{2}c) \sqrt{\hat{V}} \sin(\case{1}{2}c)\right)^2 \chi_j &=&
 \left(\sqrt{V_j}- \case{1}{2} \sqrt{V_{j+\frac{1}{2}}}- \case{1}{2}
     \sqrt{V_{j-\frac{1}{2}}} \right)^2 \chi_j\\
 \left(\sin(\case{1}{2}c) \sqrt{\hat{V}}
   \cos(\case{1}{2}c)-\cos(\case{1}{2}c) \sqrt{\hat{V}}
   \sin(\case{1}{2}c)\right)^2\chi_j &=& -\case{1}{4}\left(
   \sqrt{V_{j+\frac{1}{2}}}- \sqrt{V_{j-\frac{1}{2}}}\right)^2 \chi_j
\end{eqnarray*}
and analogously on $\zeta_j$ (the state $\zeta_{-\frac{1}{2}}$ is
annihilated by both operators, so we have to define $V_{-1}=0$ when
using the above formulae in this case).

Inserting these operators in (\ref{miso}) we immediately obtain the
eigenvalues
\begin{equation}\label{eigenm}
 m_{IJ,j}=16 (\gamma\lP^2)^{-2} \left(\delta_{IJ} \left(
   \sqrt{V_{j+\frac{1}{2}}}- \sqrt{V_{j-\frac{1}{2}}}\right)^2+
   4\left(\sqrt{V_j}- 
     \case{1}{2} \sqrt{V_{j+\frac{1}{2}}}- \case{1}{2}
     \sqrt{V_{j-\frac{1}{2}}} \right)^2\right)
\end{equation}
in terms of the volume eigenvalues. Note that there are also
non-vanishing off-diagonal terms which are independent of $I,J$. Using
(\ref{vol}) we can expand the eigenvalues of the product
$\hat{V}^{\frac{1}{3}}\hat{m}_{IJ}$, which classically should be
$\delta_{IJ}$:
\begin{eqnarray}
 m_{IJ,j} &=& V_j^{-\frac{1}{3}}\left(\delta_{IJ}+(\case{1}{256}+
  \case{37}{192}\delta_{IJ}) j^{-2}+O(j^{-3})\right)\label{expand}\\
 &\sim& V^{-\frac{1}{3}}
 \left(\delta_{IJ}+ \case{1}{9} \left(\case{1}{256}+ \case{37}{192}
     \delta_{IJ} \right) \gamma^2\lP^4 a^{-4}\right)\,.\nonumber
\end{eqnarray}
For large $j$ we used the approximation $a^2= |p| \sim
\frac{1}{3}\gamma\lP^2 j$.  This demonstrates that the leading order
is in fact given by $V^{-\frac{1}{3}} \delta_{IJ}$ and higher order
corrections only start with $\lP^4a^{-4}$. Also the {\em off-diagonal
terms}, which are of purely quantum origin, {\em only arise at this
order}. Thus, we have the correct behavior in the classical regime of
large scale factor (since the only way to obtain dimensionful
geometric quantities is by multiplying a given function of $j$ by
$\lP$, the classical limit involves with $\lP\to0$ also $j\to\infty$,
as in the treatment of angular momentum in quantum mechanics) and we
see that all the techniques involved in the quantization of the
inverse scale factor are perfectly compatible with the classical
limit. Note also that the leading order in the expansion is
independent of the Barbero--Immirzi parameter $\gamma$.

\begin{figure}[ht]
\begin{center}
 \includegraphics[width=12cm,height=8cm,keepaspectratio]{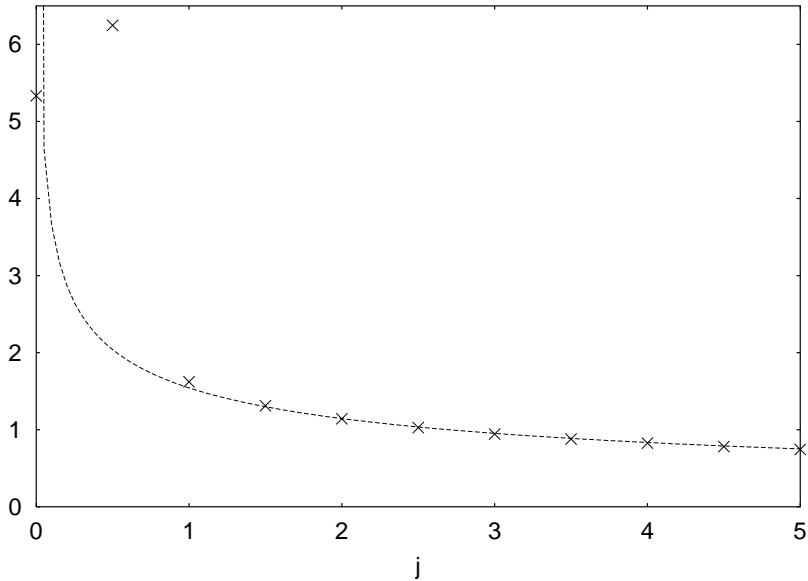}
\end{center}
\caption{Eigenvalues $m_{II,j}$, $j\geq0$
(in units of $(\gamma\lP^2)^{-\frac{1}{2}}$)
of the inverse scale factor ($\times$) compared to the classical
expectation $V_j^{-\frac{1}{3}}$ (dashed
line).}
\label{inv}
\end{figure}

\begin{figure}[ht]
\begin{center}
 \includegraphics[width=12cm,height=8cm,keepaspectratio]{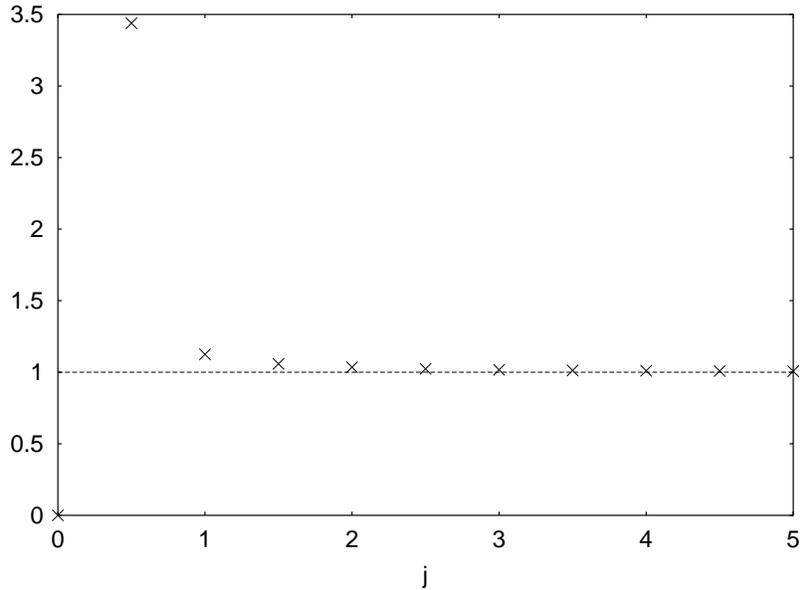}
\end{center}
\caption{Product $V_j^{\frac{1}{3}}m_{II,j}$ of eigenvalues of the
scale factor and the inverse scale factor compared to the classical
expectation one (dashed line).}
\label{prod}
\end{figure}

In fact, the classical behavior can be observed in a range by far
larger than expected from the $j^{-1}$-expansion. As can be seen in
Figs.\ \ref{inv} and \ref{prod}, even down to $j=1$, i.e.\ for a scale
factor slightly above the Planck length, lie the eigenvalues close to
the classical expectation. Only the lowest eigenvalues deviate
strongly from the classical curve, but this is in a regime where
quantum effects are important. Those effects are responsible for the
boundedness of the quantized inverse scale factor and its finite
eigenvalues even on the states $\chi_0$, $\zeta_0$ and
$\zeta_{-\frac{1}{2}}$ which are annihilated by the volume operator.

As expected from the fact that both classical quantities only depend
on the triad degrees of freedom, the volume operator $\hat{V}$ and the
quantized inverse scale factor $\hat{m}_{IJ}$ are simultaneously
diagonalizable. The surprising fact is that $\hat{m}_{IJ}$ is a
densely defined operator with the correct classical behavior in a wide
range. Thus, even in states which are annihilated by the volume
operator we must have finite eigenvalues of $\hat{m}_{IJ}$; otherwise
$\hat{m}_{IJ}$ would not be densely defined. Since $\hat{m}_{II}$ is a
quantization of the inverse scale factor, such a behavior would be
impossible in the classical description where we have the identity
$V^{\frac{1}{3}}\cdot m_{IJ}=\delta_{IJ}$. In quantum theory, while
this relation {\em is\/} valid at large volume (see Fig. \ref{prod}),
quantum corrections cause large deviations at the Planck scale close
to the zero volume states. Thus, $\hat{V}$ and $\hat{m}_{II}$ are not
inverse operators of each other, which in particular allows finite
eigenvalues of $\hat{m}_{II}$ in states annihilated by $\hat{V}$. This
is a new mechanism with origin purely in quantum geometry by which the
classical singularity is resolved.

\subsection{The $j=-\frac{1}{2}$-state}
\label{s:Half}

Two aspects of the quantization of the inverse scale factor are
important: first, it is a bounded operator cutting off the classical
divergence and simultaneously preserving the classical behavior for
values of the volume larger (not much larger) than a Planck cube. This
fact is welcome and can be understood as originating in quantum
effects which become important only for small volume where the
classical theory breaks down. Technically, this is done by choosing an
appropriate classical expression as the starting point for
quantization. Since the volume operator has zero eigenvalues, its
inverse does not exist and can, therefore, not be used for a
quantization of the inverse scale factor. But as we have seen, it is
possible to rewrite the inverse scale factor as the expression
(\ref{m}) which is quantized to a bounded operator (\ref{miso}). In
this way, it is understandable that the quantization $\hat{m}_{IJ}$ of
the classical quantity $m_{IJ}=|a|^{-1}\delta_{IJ}$ does not coincide
with the (non-existing) inverse ``$\hat{a}^{-1}$'' of the quantization
of $a$. Such effects are not unexpected in quantum theory. A second
feature of the quantization of the inverse scale factor seems to be
more questionable: one of the zero volume eigenstates has also zero
(not just finite) eigenvalue of the inverse scale factor and so both
$\hat{a}$ and $\hat{m}_{II}$ annihilate the same state. Of course,
this happens at a point where the classical theory completely breaks
down and classical intuition cannot be trusted, but still an
elucidation is needed\footnote{The author is grateful to A.~Ashtekar
for raising this issue.}. This is even more important because these
issues are essential for a quantum evolution through the classical
singularity \cite{Sing,IsoCosmo}.

A basic observation in this respect is that the classical value of the
inverse scale factor $m_{IJ}$ corresponding to the quantization
(\ref{mhom}) is not defined at $a=0$, and so $m_{IJ}$ has to be
appropriately extended to this point. This can formally be done as
$m_{IJ}:= \sgn(a)^2|a|^{-1}\delta_{IJ}$ (taking into account the
derivative of $V=|p|^{\frac{3}{2}}$ in the Poisson bracket) which, of
course, is the same as (\ref{m}) on the classically allowed region
$a>0$. In the point $a=0$ both this expression and $a^{-1}$ are
ill-defined and so its introduction does not change the classical
situation. (The sign does not lead to a better behavior since it is,
as the derivative of the absolute value, not well-defined for $a=0$;
and even the standard definition $\sgn(0):=0$ leads to an undefined
expression ``$0/0$'', whereas the limit $a\to 0$ is not different from
the one for $a^{-1}$.) However, as we have seen the situation is very
different upon quantization which leads to a well-defined formulation
also at states corresponding to the classical value $a=0$.

In order to illustrate this point, one can do essentially the some
quantization in a simpler model which is the standard quantum theory
of the cylinder $T^*S^1$ with canonically conjugate coordinates
$\{\phi,\pi\}=1$. Its states are $|n\rangle=\exp(in\phi)$ on which the
basic operators act as
\begin{eqnarray}
 \cos\phi\,|n\rangle &=&
 \case{1}{2}(|n+1\rangle+|n-1\rangle)\label{cosS1}\\
 \sin\phi\,|n\rangle &=&
 -\case{1}{2}i(|n+1\rangle-|n-1\rangle)\label{sinS1}
\end{eqnarray}
and
\begin{equation}
 \hat{\pi}|n\rangle=n\hbar|n\rangle\,. \label{p}
\end{equation}
Being interested in a quantization of $|\pi|^{-\frac{1}{2}}$, one
cannot use the inverse of $\hat{\pi}$ which does not exist. Instead,
one can rewrite
\begin{equation}\label{pinv}
 {\rm sgn}(\pi)|\pi|^{-\frac{1}{2}}=2\left(\cos\phi
 \left\{\sin\phi,\sqrt{|\pi|}\right\}-
 \sin\phi\left\{\cos\phi,\sqrt{|\pi|}\right\}\right)
\end{equation}
which is a classical identity and clearly shows the origin of the
sign. We used a similar identity (Thiemann's) to rewrite the inverse
scale factor in isotropic cosmological models. It is also clear that
one cannot use an analogous formula to rewrite $|\pi|^{-\frac{1}{2}}$
itself, since one always needs to take derivatives of the absolute
value of $\pi$. But in $\pi=0$ both classical expressions are
ill-defined and so there is no ``correct'' one as a starting point for
quantization. The only difference is that, as we will see shortly,
$\sgn(\pi)|\pi|^{-\frac{1}{2}}$ can be quantized to a densely
defined operator, whereas $|\pi|^{-\frac{1}{2}}$ cannot.

Expression (\ref{pinv}) can easily be quantized by turning the Poisson
brackets into $(i\hbar)^{-1}$ times commutators which leads to
\begin{equation}
 \sgn(\hat{\pi})|\hat{\pi}|^{-\frac{1}{2}}=
 -2i\hbar^{-1}\left(\cos\phi
 \left[\sin\phi,\sqrt{|\hat{\pi}|}\right]-
 \sin\phi\left[\cos\phi,
\sqrt{|\hat{\pi}|}\,\right]\right)\,.
\end{equation}
Using (\ref{cosS1}), (\ref{sinS1}) and (\ref{p}), its action on the states
$|n\rangle$ can be computed which shows that $|n\rangle$ are
eigenstates with eigenvalues
\begin{equation}
 \sqrt{|n+1|\hbar^{-1}}-\sqrt{|n-1|\hbar^{-1}}\,.
\end{equation}
For large $|n|$, one can perform a Taylor expansion demonstrating the
correct classical limit
\begin{eqnarray*}
 \sqrt{|n+1|\hbar^{-1}}-\sqrt{|n-1|\hbar^{-1}} &=&
 \hbar^{-\frac{1}{2}} \sqrt{|n|}\left(\sqrt{1+n^{-1}}-
 \sqrt{1-n^{-1}}\right)\\
 &=& \hbar^{-\frac{1}{2}} \sqrt{|n|} \left(n^{-1}+O(n^{-3})\right)\\
 &=& {\rm sgn}(n)(|n|\hbar)^{-\frac{1}{2}}(1+O(n^{-2}))\,.
\end{eqnarray*}
On the other hand, for $n=0$ the eigenvalue is zero, which is the same
situation as in the quantization of the inverse scale factor: both
$\hat{\pi}$ and $\sgn(\hat{\pi})|\hat{\pi}|^{-\frac{1}{2}}$ annihilate
the same state. In view of the sign appearing with the inverse this is
less surprising than initially, but again we emphasize that the sign
makes no difference for the acceptability of the classical
quantity. This toy example demonstrates that the important features of
the quantization of the inverse scale factor are not special and can
also be obtained in standard quantum mechanics. Nevertheless one may
ask why such quantizations have not been used before. The answer is
related to the classically allowed regions of the canonical
coordinates, which are different in the $T^*S^1$-example and in
isotropic cosmological models: on the cylinder, the full range for
$\pi$ is allowed including $\pi=0$ and so the inverse of $\pi$ is not
well-defined and also not of physical interest. But in the gravity
model, the scale factor $a$ must be positive classically, and so
$a^{-1}$ is well-defined on the classical phase space. In contrast to
the inverse angular momentum $\pi^{-1}$, it is of direct physical
interest since it appears, e.g., in curvatures.

\section{Discussion}

Isotropic cosmological models are not only interesting as models of an
expanding universe, for which they have been widely used in the past
century, but also provide systems in which explicit calculations are
feasible, and therefore are excellent test arenas for sophisticated
techniques developed for a quantization of general relativity. The
basic technique used in this paper is the quantization of the co-triad
components due to Thiemann \cite{QSDI}. Most of the operators studied
so far in quantum geometry are built from the fundamental holonomies
or derivative operators quantizing the triad components. The co-triad,
however, is not a fundamental field in this framework, and so its
quantization is less clear-cut. Nevertheless, it is of importance because
it enters the quantization of the Hamiltonian constraint, which
controls the dynamics of the theory. Models like the isotropic ones
can be used to investigate quantization ambiguities in detail, and we
presented a first test of the co-triad quantization by studying a
quantization of the inverse scale factor which can be expressed in
terms of the co-triad. As we have seen, applying Thiemann's identity
leads to a bounded operator while preserving the correct classical
limit. Although quantization ambiguities like factor ordering problems
do not affect the classical limit, the outcome of the correct
classical limit is not trivial. A look at the eigenvalues
(\ref{eigenm}) reveals that a large-$j$ behavior $V_j\sim
j^{\frac{d}{2}}$ leads to $m_{II,j}\sim j^{-\frac{d}{6}}\sim
V_j^{-\frac{1}{3}}$, which is needed for the correct classical limit
of the inverse scale factor, only for $d=3$. Also the prefactor is
important because otherwise the product $V_j^{\frac{1}{3}}m_{II,j}$
would not approach one for large $j$. For this it was necessary to use
the correct symplectic structure (\ref{symp}) and also to correct the
volume spectrum (\ref{vol}). The quantization techniques of
\cite{QSDI} are working perfectly in our quantization which should
also increase our confidence in the quantization of the Hamiltonian
constraint in the full theory.

An expansion for large $j$ showed the correct classical behavior
(\ref{expand}); but the situation is even better than expected: such
an expansion is supposed to break down for $j$ being of the order one,
but Fig.\ \ref{inv} shows that the eigenvalues of the inverse scale
factor are close to the classical expectation even down to $j=1$. In
fact, there is a cancelation in the $j^{-1}$-correction to the
classical expression in (\ref{expand}), but this still cannot explain
the good behavior for $j=1$: the Taylor expansion of (\ref{eigenm})
around $x=0$ for $x:=j^{-1}$, which has been used in (\ref{expand}),
does not converge for $x=1$ corresponding to the
$j=1$-eigenvalue. This fact shows that a non-perturbative treatment is
mandatory: a perturbative expansion in $j^{-1}$, which is meaningful
for large scale factors, breaks down at the Planck scale.

Close to the classical singularity, quantization ambiguities do affect
the eigenvalues quantitatively, but not the qualitative conclusion of
a non-diverging behavior of the inverse scale factor. One can draw
lessons from our quantization, e.g., one could replace the objects
$-h[h^{-1},\hat{V}]$ by $[h,\hat{V}]$ in isotropic models because all
edges are closed there and both correspond to the same classical
Poisson bracket (this has been done in \cite{cosmoIII} for
simplicity). In the full theory, this is not possible because any edge
appearing in a holonomy has to be traced back. One can see that such a
replacement would not lead to the desired properties for the inverse
scale factor in isotropic models because the resulting expression
would not commute with the volume operator (what it should do because
both only depend on metrical variables). Thus, one has to use the same
procedure as in the full theory despite of a larger initial freedom.

\begin{figure}[ht]
\begin{center}
 \includegraphics[width=12cm,height=8cm,keepaspectratio]{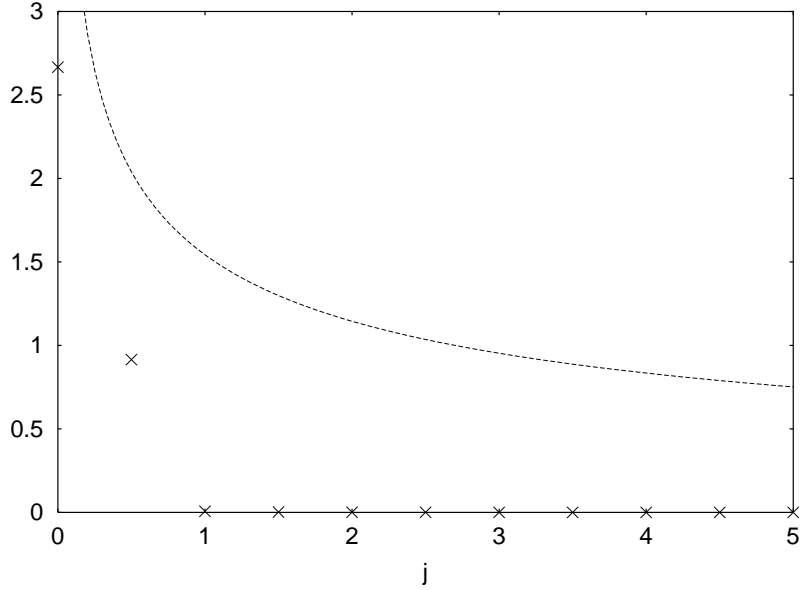}
\end{center}
\caption{Eigenvalues of the off-diagonal components of $\hat{m}_{IJ}$
compared to the classical inverse scale factor (dashed line).}
\label{nond}
\end{figure}

A consequence of the particular quantization presented in Section
\ref{s:Inv} is the fact that the metric eigenvalues form a
non-diagonal matrix even for an isotropic model. The off-diagonal
terms are, however, only of the order $\lP^4/a^4$ and so do not affect
the classical limit (see Fig.\ \ref{nond}). Their precise value
depends on the factor ordering and other ambiguities which cannot be
fixed by studying the classical limit. For example, they can be
removed by quantizing the product of the co-triads by
\begin{eqnarray}\label{trtr}
 &&\tr\left(\tau_i h_I\left[h_I^{-1},\sqrt{\hat{V}}\right]\right)
 \tr\left(\tau_i h_J \left[h_J^{-1}, \sqrt{\hat{V}}\right]\right)
 = \case{1}{2} \tr\left(h_I\left[h_I^{-1},\sqrt{\hat{V}}\right]
 \left(h_J\left[h_J^{-1},\sqrt{\hat{V}}\right]\right)^{-1}\right)\nonumber\\
 &&\quad- \case{1}{4}
 \tr\left(h_I\left[h_I^{-1},\sqrt{\hat{V}}\right]\right)
 \tr\left(h_J\left[h_J^{-1},\sqrt{\hat{V}}\right]\right)
\end{eqnarray}
instead of (\ref{mhom}). Here we used the identity $(\tau_i)^A{}_B
(\tau_i)^C{}_D= \frac{1}{2}\epsilon^{AC}\epsilon_{BD}-
\frac{1}{4}\delta^A{}_B\delta^C{}_D$ and defined
$(h^{-1})^A{}_B:=\epsilon^{AC}\epsilon_{BD} h^D{}_C$. The latter is an
identity for the inverse of $h$ if $h\in SU(2)$, but the commutator
$h_J[h_J^{-1},\sqrt{\hat{V}}]$ is in general not invertible as an
(operator valued) $2\times2$-matrix. Therefore,
$(h_J[h_J^{-1},\sqrt{\hat{V}}])^{-1}$ is just a short form for
\begin{eqnarray*}
 {\left(\left(h_J[h_J^{-1},\sqrt{\hat{V}}]\right)^{-1}\right)^A}_B
 &:=& \epsilon^{AC}\epsilon_{BD} {\left(h_J[h_J^{-1},
 \sqrt{\hat{V}}]\right)^D}_C
 =(h_J^{-1})^F{}_B \left[(h_J)^A{}_F,\sqrt{\hat{V}}\right]\\
 &=& \left(\sqrt{\hat{V}}-
 \cos(\case{1}{2}c)\sqrt{\hat{V}}\cos(\case{1}{2}c)-
 \sin(\case{1}{2}c)\sqrt{\hat{V}}\sin(\case{1}{2}c)\right)\delta^A{}_B\\
 && +2\Lambda^A{}_B \left(\sin(\case{1}{2}c)\sqrt{\hat{V}}\cos(\case{1}{2}c)-
 \cos(\case{1}{2}c)\sqrt{\hat{V}}\sin(\case{1}{2}c)\right).
\end{eqnarray*}
The subtraction of the product of traces in (\ref{trtr}) leads to a
cancellation of the off-diagonal components in the resulting
$\hat{m}_{IJ}$.  However, there is no independent argument in favor of
(\ref{mhom}) or (\ref{trtr}) besides the vanishing of off-diagonal
components since both expression have the same classical
limit. Off-diagonal components may have relevance in deviations from
the Lorentz-invariant vacuum structure because a non-diagonal metric
at small scales leads to anisotropies and so to birefringence
effects in the propagation of waves. In this context, it may be
interesting that the corrections are only of fourth order in the
Planck length.

The main result of this paper is that the divergence of the inverse
scale factor is completely cured by the quantization methods of
quantum geometry, most importantly those developed in \cite{QSDI}.
This fact opens up a new way for a resolution of the classical
singularity in quantum cosmology \cite{Sing} which will be
investigated in more detail elsewhere \cite{IsoCosmo}. Technically,
this comes from an absorption of $V^{-1}$ into a Poisson bracket which
is the same procedure which allows matter Hamiltonians to be quantized
to densely defined operators \cite{QSDV}.  Therefore, {\em it is the
same mechanism which regularizes ultraviolet divergencies in quantum
field theories and removes the classical singularity in quantum
cosmology}. In particular, geometry itself is responsible for this to
happen, and not matter effects.

\section*{Acknowledgements}

The author is grateful to A.\ Ashtekar for discussions and a careful
reading of the manuscript.  This work was supported in part by NSF
grant PHY00-90091 and the Eberly research funds of Penn State.


\begin{thebibliography}{10}

\bibitem{DeWitt}
B.~S.\ DeWitt,
\newblock Quantum Theory of Gravity. I. The Canonical Theory,
\newblock {\em Phys.\ Rev.} 160 (1967) 1113--1148

\bibitem{Misner}
C.~W.\ Misner,
\newblock Quantum Cosmology. I,
\newblock {\em Phys.\ Rev.} 186 (1969) 1319--1327

\bibitem{Rov:Loops}
C.\ Rovelli,
\newblock Loop Quantum Gravity,
\newblock {\em Living Reviews in Relativity} 1 (1998)
  http://www.livingreviews.org/Articles/Volume1/1998--1rovelli, [gr-qc/9710008]

\bibitem{AreaVol}
C.\ Rovelli and L.\ Smolin,
\newblock Discreteness of Area and Volume in Quantum Gravity,
\newblock {\em Nucl.\ Phys.\ B} 442 (1995) 593--619, [gr-qc/9411005],
\newblock Erratum: {\em Nucl.\ Phys.\ B} 456 (1995) 753

\bibitem{Vol2}
A.\ Ashtekar and J.\ Lewandowski,
\newblock Quantum Theory of Geometry II: Volume Operators,
\newblock {\em Adv.\ Theor.\ Math.\ Phys.} 1 (1997) 388--429, [gr-qc/9711031]

\bibitem{SymmRed}
M.\ Bojowald and H.~A.\ Kastrup,
\newblock Symmetry Reduction for Quantized Diffeomorphism Invariant Theories of
  Connections,
\newblock {\em Class.\ Quantum Grav.} 17 (2000) 3009--3043, [hep-th/9907042]

\bibitem{cosmoI}
M.\ Bojowald,
\newblock Loop Quantum Cosmology: I. Kinematics,
\newblock {\em Class.\ Quantum Grav.} 17 (2000) 1489--1508, [gr-qc/9910103]

\bibitem{cosmoII}
M.\ Bojowald,
\newblock Loop Quantum Cosmology: II. Volume Operators,
\newblock {\em Class.\ Quantum Grav.} 17 (2000) 1509--1526, [gr-qc/9910104]

\bibitem{cosmoIII}
M.\ Bojowald,
\newblock Loop Quantum Cosmology III: Wheeler-DeWitt Operators,
\newblock {\em Class.\ Quantum Grav.} 18 (2001) 1055--1070, [gr-qc/0008052]

\bibitem{cosmoIV}
M.\ Bojowald,
\newblock Loop Quantum Cosmology IV: Discrete Time Evolution,
\newblock {\em Class.\ Quantum Grav.} 18 (2001) 1071--1088, [gr-qc/0008053]

\bibitem{PhD}
M.\ Bojowald,
\newblock {\em Quantum Geometry and Symmetry},
\newblock PhD thesis, RWTH Aachen, 2000,
\newblock published by Shaker-Verlag, Aachen

\bibitem{QSDI}
T.\ Thiemann,
\newblock Quantum Spin Dynamics {(QSD)},
\newblock {\em Class.\ Quantum Grav.} 15 (1998) 839--873, [gr-qc/9606089]

\bibitem{QSDV}
T.\ Thiemann,
\newblock {QSD V}: Quantum Gravity as the Natural Regulator of Matter Quantum
  Field Theories,
\newblock {\em Class.\ Quantum Grav.} 15 (1998) 1281--1314, [gr-qc/9705019]

\bibitem{FermionHiggs}
T.\ Thiemann,
\newblock Kinematical Hilbert Spaces for Fermionic and Higgs Quantum Field
  Theories,
\newblock {\em Class.\ Quantum Grav.} 15 (1998) 1487--1512, [gr-qc/9705021]

\bibitem{Sing}
M.\ Bojowald,
\newblock Absence of a Singularity in Loop Quantum Cosmology,
\newblock {\em Phys.\ Rev.\ Lett.} (2001) in press, [gr-qc/0102069]

\bibitem{IsoCosmo}
M.\ Bojowald,
\newblock Isotropic Loop Quantum Cosmology,
\newblock in preparation

\end{thebibliography}
\end{document}